\documentclass[a4paper,12pt]{article}
\title{\textbf{MICROSCOPIC SPACE-TIME OF EXTENDED HADRONS :
QUANTUM GENERALIZATION AND FIELD EQUATIONS}}
\date{}
\usepackage{amssymb}
\usepackage{latexsym}
\usepackage{graphicx}
\begin{document}
\setcounter{section}{0}
\renewcommand{\thesection}{\Roman{section}.}
\setcounter{equation}{0}
\renewcommand{\theequation}{\arabic{equation}}
\maketitle
\begin{center}
S.S.DE \\
\textit{Department of Applied Mathematics, Calcutta
University,
92,A.P.C. Road, Kolkata - 700009, INDIA}\\
e-mail : satya@cubmb.ernet.in\\
\end{center}
\begin{abstract}
The microlocal space-time of extended hadrons, considered to be
anisotropic is specified here as a special Finsler space. For this
space the classical field equation is obtained from a property of
the field on the neighbouring points of the autoparallel curve.
The quantum field equation has also been derived for the bispinor
field of a free lepton in this Finslerian microspace through its
quantum generalization below a fundamental length-scale. The
bispinor can be decomposed as a direct product of two spinsors,
one depending on the position coordinates and the other on the
directional arguments of the Finsler space. The former one
represents the spinor of the macrospace, an associated Riemannian
space-time of the Finsler space, and satisfies the Dirac equation.
The directional variable-dependent spinor satisfies a different
equation which is solved here. This spinor-part of the bispinor
field for a constituent of the hadron can give rise to an
additional quantum number for generating the internal symmetry of
hadrons. Also, it is seen that in the process of separating the
bispinor field and its equation an epoch-dependent mass term
arises. Although, this part of the particle-mass has no
appreciable contribution in the present era it was very
significant for the very early period of the universe after its
creation. Finally, the field equations for a particle in an
external electromagnetic field for this Finslerian microlocal
space-time and its associated Riemannian macrospaces have been
found.
\end{abstract}

\noindent
\section {INTRODUCTION}

     As early as in 1957, the measurement of
electromagnetic form factor of nucleon may be regarded as the
first experimental support for the concept of extended structure
of the elementary particle (Hofstadter, 1964), although,
theoretically it was an older idea. In fact, this concept can be
traced back in the electron theory of Lorentz. On the contrary,
the elementary particles as the field quanta which are essentially
point-like entities originated from the relativistic quantum
mechanics in the framework of local field theory which suffers
from the well-known divergence difficulties. With an underlying
motivation of resolving these difficulties and also for the
unified description of elementary particles, Yukawa (1948, 1950)
came up with a bilocal field theory which opened the possibility
of intrinsic extensions of the elementary particles. Later this
simple bilocal model was generalized to multilocal theories. The
motivation behind all these theories lies in understanding
multifariousness of subatomic particles (whose number went on
increasing in the last fifty years) as the ultimate manifestation
of their extensions in space and time, the microscopic space-time.

    The hadronic matter-extension in the microlocal space-time is also
manifested in the composite character of hadrons, and the
investigations of such composite pictures of extended hadrons
culminated into the theory of quarks by Gell-Mann (1964), and the
quark-parton model of Feynman (Bjorken, 1969; Feynman, 1972). So
far, no direct observation of free quarks is possible, and
consequently the confinement of quarks in the structure of hadrons
has to be assumed. Thus, in a sense, these quarks can not be
considered as the ordinary particles. Moreover, not only their
fractional charges but also their flavour quantum numbers, such as
isospin, strangeness, charm, etc. have to be assigned for
generating the quantum numbers of hadrons. Even in Weinberg-Salam
unified theory of weak and electromagnetic interactions such a
phenomenology of assigning individual hypercharges to leptons and
quarks has to be adopted. Since the origin of these internal
quantum numbers of the constituents is not found, the
quark-constituent model of extended hadron-structure remains
incomplete. On the other hand, if one considers the extensions of
hadron-structure in the microlocal space-time, then there remains
the problem of specification of this space-time, which may be
different from the macroscopic spaces, such as, the laboratory
(Minkowskian) space-time and the large-scale space-time of the
universe. Apart from the specification of the microscopic
space-time, one has to obtain the field equations for this
space-time with an aim of constructing a field theory of hadron
interactions. The geometric origin of internal quantum numbers of
the constituents is also to be found, making the internal symmetry
of hadrons possible.

     Another important aspect one has to remember
is that the space-time below a length-scale (the Planck scale) is
not a meaningful concept. Recently, Adler and Santiago (1999) have
modified the uncertainty principle by considering the
gravitational interaction of the photon and the particle. From
this modified gravitational uncertainty principle it follows that
there is an absolute minimum uncertainty in the position of any
particle and it is of the order of Planck length. This is also a
standard result of superstring theory. Also, the intrinsic
limitation to quantum measurements of space-time distances has
been obtained (Ng and Van Dam, 1994; and references therein). This
uncertainty in space-time measurements does also imply an
intrinsic uncertainty of the space-time metric and yields a
quantum decoherence for particles heavier than Planck mass. Thus,
because of the intrinsic uncertainty of space-time metric, it
suffices to give a particle heavier than the Planck mass a
classical treatment. Also, the space-time can be defined only as
averages over local regions and cannot have any meaning locally.
This also indicates that one should treat the space-time as
"quantized" below a fundamental length-scale.

     In the present article, the microlocal space-time is regarded as a
special
Finsler space which is anisotropic in nature. In fact, the
breaking of discrete space-time symmetries in weak interactions,
an anisotropy in the relic background radiation of the universe,
and the absence of the effect of cutoff in the spectrum of primary
ultra-high energy cosmic protons are all indirect indications of
the existence of a local anisotropy in space-time. Consequently,
the microlocal space-time should be described by Finsler geometry
instead of Riemannian one. Here, we begin, in section 2, with the
construction of the classical field equation in the Finsler space.
In section 3, a discussion on the quantum generalization of the
space-time has been made. In the subsequent section 4, the
microlocal space-time has been considered as a quantized special
Finsler space; and in section 5, the field equation for this
Finsler space of microdomain has been derived. The field for a
free lepton is, here, a "bispinor" depending on the position
coordinates and the directional variables, and from its field
equation it is possible to find the Dirac equations for the
spinors of the macrospaces, such as, the usual Minkowski
space-time and the background space-time of the universe (the
Robertson-Walker space-time). These macrospaces appear as the
associated Riemannian spaces of the Finslerian microscopic space.
Of course, an alternative approach of deriving the Dirac equations
for macrospaces has also been suggested in section 6. In the next
section 7, we have found the homogeneous solutions of the
directional variable-dependent spinor-part from the separable
bispinor field. This spinor can give rise to an additional quantum
number for the constituent leptons for generation of the internal
symmetry of hadrons. Also, in the process of decomposition of the
bispinor an additional mass term depending on the cosmological
time appears. The cosmological consequences of this mass term were
considered elsewhere (De, 1999; and the references therein). In
section 8, the field equations for a particle in an external
electromagnetic field have been obtained for both the microscopic
and macroscopic space-times. In the final section 9, some
concluding remarks have been made.

\section {CLASSICAL FIELD EQUATION}

     The classical field $\psi(\textbf{x},\mathbf{\nu})$
corresponding to a free particle supposed to be  a lepton depends
on the directional variables
$\mathbf{\nu}=(\nu^0,\nu^1,\nu^2,\nu^3)$ of the Finsler space
apart from its dependence  on the position coordinates
$\textbf{x}=({x}^0,{x}^1,{x}^2,{x}^3)$ as this field is an
entity of the Finsler space. In fact, all the geometrical
objects of any Finsler space, such as the metric tensor, the
connection coefficients, depend on the positional as well as
directional arguments. The classical field equation for
$\psi(\textbf{x},\mathbf{\nu})$ can be obtained if one admits the
following conjecture (De, 1997) : A property is to be satisfied by
$\psi(\textbf{x},\mathbf{\nu})$ along the neighbouring points in
the Finslerian microdomain on the autoparallel curve (the geodesic
) which is the curve whose tangent vectors result from each other
by successive infinitesimal parallel displacement of the type
\begin{equation}
d\nu^i=-\gamma_{hj}^i(\textbf{x},\mathbf{\nu})\nu^h dx^j
\end{equation}
where $\gamma_{hj}^i(\textbf{x},\mathbf{\nu})$ are the Christoffel
symbols of second kind. Now, the property can be stated as the
infinitesimal change of $\psi$ along the autoparallel curve is
proportional to the field function itself, that is,
 $$\delta\psi=\psi(\textbf{x}+d\textbf{x},\mathbf{\nu}+d\mathbf{\nu})
 -\psi(\textbf{x},\mathbf{\nu})\propto
 \psi(\textbf{x},\mathbf{\nu})$$
or,
\begin{equation}
\psi(\textbf{x}+d\textbf{x},\mathbf{\nu}+d\mathbf{\nu})
 -\psi(\textbf{x},\mathbf{\nu})=\epsilon (mc/\hbar)
\psi(\textbf{x},\mathbf{\nu})
\end{equation}
where the neighbouring points $(
\textbf{x}+d\textbf{x},\mathbf{\nu}+d\mathbf{\nu})$ and
$(\textbf{x},\mathbf{\nu})$ lie on the autoparallel curve. Here,
the mass term m appearing in the constant of proportionality is
regarded as the inherent mass of the particle and $\epsilon$ is
a real parameter such that$ 0 < \epsilon \leq \ell, \ell$ being a
fundamental length.

    To the first orders in $d\textbf{x} and d\mathbf{\nu}$ we have, from
(2),

\begin{equation}
(dx^\mu \partial_\mu + d\nu^\ell
\partial_\ell^\prime)\psi(\textbf{x},\mathbf{\nu})=\epsilon (mc/\hbar)
\psi(\textbf{x},\mathbf{\nu})
\end{equation}

where $\partial_\mu \equiv \partial/\partial x^\mu and
\partial_\ell^\prime = \partial/\partial\nu^\ell$

Using (1), we arrive at the following equation for
$\psi(\textbf{x},\mathbf{\nu})$ :
\begin{equation}
dx^\mu(\partial_\mu - \gamma_{h\mu}^\ell
(\textbf{x},\mathbf{\nu})\nu_h
\partial_\ell^\prime)\psi(\textbf{x},\mathbf{\nu}) = \epsilon (mc/\hbar)
\psi(\textbf{x},\mathbf{\nu})
\end{equation}

This equation can also be written in terms of the nonlinear
connection $(N_\mu^\nu)$ (in local representation) given as
\begin{equation}
N_\mu^\nu =(1/2)
\frac{\partial}{\partial\nu^\nu}(\gamma_{\alpha\beta}^\mu
\nu^\alpha \nu^\beta )
\end{equation}
 Because of homogeneity property, one finds
 $(N_\mu^\nu)\nu^\nu =\gamma_{\alpha\beta}^\mu \nu^\alpha
\nu^\beta )$ where $\nu^\alpha = \frac{dx^\alpha}{ds}.$
Consequently, we have
\begin{equation}
N_\mu^\nu dx^\nu =\gamma_{\alpha\beta}^\mu \nu^\alpha dx^\beta
\end{equation}
Then, the equation (4) becomes
\begin{equation}
dx^\mu (\partial_\mu - N_\mu^\ell \partial_\ell^\prime)
\psi(\textbf{x},\mathbf{\nu})=\epsilon (mc/\hbar)
\psi(\textbf{x},\mathbf{\nu})
\end{equation}
or, in terms of covariant   $\frac{\delta}{\delta x^\mu}
\equiv\partial_\mu - N_\mu^\ell \partial_\ell^\prime$ we have
\begin{equation}
dx^\mu \frac{\delta\psi(\textbf{x},\mathbf{\nu})}{\delta x^\mu}
=\frac{\epsilon mc}{\hbar} \psi(\textbf{x},\mathbf{\nu})
\end{equation}
Now, as $dx^\mu$ transforms covariantly like a vector,
$\psi(\textbf{x},\mathbf{\nu})$ behaves like a scalar under the
general coordinate transformations, that is,
\begin{equation}
\psi^\prime(\textbf{x}^\prime,\mathbf{\nu}^\prime)=\psi(\textbf{x},\mathbf{\nu})
\end{equation}
where  $x^{\prime\mu}=x^{\prime\mu}(x^\mu)$ and $\nu^{\prime\mu}=
X_\nu^{*\mu} \nu^\nu$
 with   $X_\nu^{*\mu}=\frac{\partial x^{\prime\mu}}{\partial x^\nu}$

It is to be noted that the equation (8) for the classical field
function is form-invariant under the general coordinate
transformations. Now, if the parameter $\epsilon$ is identified
with the Finslerian arc distance element ds between the
neighbouring points on the autoparallel curve, that is, the
geodesic distance element, then we get the classical field
equation in the following form :
\begin{equation}
 \nu^\mu \frac{\delta\psi(\textbf{x},\mathbf{\nu})}{\delta x^\mu}
=\frac{mc}{\hbar} \psi(\textbf{x},\mathbf{\nu})
\end{equation}

The equations (8)  and  (10)  for the field
$\psi(\textbf{x},\mathbf{\nu})$  are derived from its property on
the autoparallel curve at the neighbouring points
$(x^\mu,\nu^\mu)$ and $(x^\mu + dx^\mu,\nu^\mu + d\nu^\mu)$ for
which $\eta_{\mu\nu} dx^\mu dx^\nu < 0$. On the contrary, for
$\eta_{\mu\nu}dx^\mu dx^\nu > 0$,this property should be taken as
\begin{equation}
\delta\psi = \psi(\textbf{x} + d\textbf{x},\mathbf{\nu} +
d\mathbf{\nu}) -\psi(\textbf{x},\mathbf{\nu})= -\frac{i\epsilon
mc}{\hbar} \psi(\textbf{x},\mathbf{\nu})
\end{equation}
which leads to the following equation for
$\psi(\textbf{x},\mathbf{\nu})$ :
\begin{equation}
dx^\mu \frac{\delta\psi(\textbf{x},\mathbf{\nu})}{\delta x^\mu}
=\frac{-i\epsilon mc}{\hbar} \psi(\textbf{x},\mathbf{\nu})
for\eta_{\mu\nu}dx^\mu dx^\nu > 0
\end{equation}

Thus, the field equation takes the following form :
\begin{equation}
i dx^\mu \frac{\delta\psi(\textbf{x},\mathbf{\nu})}{\delta x^\mu}
=\frac{\epsilon mc}{\hbar}\sqrt{\theta(d\textbf{x}^2)}
\psi(\textbf{x},\mathbf{\nu})
\end{equation}
where $d\textbf{x}^2 =\eta_{\mu\nu}dx^\mu dx^\nu$ and
\begin{equation}
\begin{array}{cc}
  \theta (z)&=1 \, \, for z\geq 0 \\
   &=-1 \,\, for z<0 \\
\end{array}\Big\}
\end{equation}
On identification of $\epsilon$  with the arc distance element ,
the above equation for $\psi(\textbf{x},\mathbf{\nu})$ becomes
\begin{equation}
i\nu^\mu \frac{\delta\psi(\textbf{x},\mathbf{\nu})}{\delta x^\mu}
=\frac{mc}{\hbar}\sqrt{\theta(\mathbf{\nu}^2)}\psi(\textbf{x},\mathbf{\nu})
\end{equation}
where $$\mathbf{\nu}^2 = \eta_{\mu\nu} \nu^\mu \nu^\nu$$

\section{QUANTIZED SPACE-TIME}
\noindent

     From the classical field functions we now transit to the quantum
field or wave functions and their equations on the basis of
quantized space-time. Snyder (1947), and Yang (1947) discussed
long ago such theories of quantized spact-time. Later, many
authors developed the theory [for various references see
Blokhintsev (1973), Prugovecki (1984) and Namsrai (1985) ]. In the
theory of quantized space-time there is no usual conceptual
meaning of definite space-time points. In fact, the components of
the operators of coordinates are not commuted. Blokhintsev (1973)
made a general statement of this problem as follows :

    The usual (c - number) coordinates of points $(x^0, x^1, x^2, x^3)$ of
space-time, which form a differential manifold
$\textbf{M}_4(\textbf{x})$ (with a certain metric) are changed by
linear operators $(\hat{x}^0, \hat{x}^1, \hat{x}^2, \hat{x}^3)$
noncommuting with one another. This leads to the question
concerning the measurable numerical coordinates of a point event
and the ordering of events in the "operational space"
$\textbf{M}_4(\hat{\textbf{x}})$ The reasonable answer to this
question can be provided by admitting a mapping of the operational
space on a space of eigenvalues of $\hat{x}$  or of functions
$f(\hat{x})$ which form a complete set of variables. This set
should be sufficient for ordering points in the four-dimensional
pseudo-Euclidean space. Toward such a consideration Blokhintsev
postulated the space $\mathcal{H}(\phi)$ of eigenfunctions $\phi$
of the complete set at each point of space
$\textbf{M}_4(\textbf{x})$. Out of the several examples considered
there the following will be relevant for the present
consideration. That was the quantum generalization of the usual
Minkowski four-dimensional space-time regarded as a special
Finsler space. The length element ds for this case is expressed as
\begin{equation}
ds = N_\mu dx^\mu
\end{equation}
where the vector $N_\mu$ is a zero-order form in $d\textbf{x}$.
This form is different for time-like,space-like directions and
light cone, having three possible values,$\textbf{N}=\pm 1,0.$

    Now, the quantum generalization of this Finsler space is achieved
through the change of coordinate differentials  in (16) by the
finite operators
\begin{equation}
\Delta x^\mu = a \gamma^\mu
\end{equation}
where $\gamma^\mu$ where gm are the Dirac matrices and a is a
certain length. The operator of interval is taken as follows :

$$\Delta \hat{s}= N_\mu \Delta \hat{x}^\mu \hspace{.5em}for
\hspace{1em} \textbf{N}^2=1 \hspace{1em}and
\hspace{1em}\textbf{N}^2=0\eqno(18a)$$
$$\Delta \hat{s}=\frac{1}{i} N_\mu \Delta \hat{x}^\mu \hspace{1em}
for \hspace{1em}  \textbf{N}^2=-1 \eqno(18b)$$

Evidently, from (17) it follows that

$$[\Delta \hat{x}^\mu, \Delta \hat{x}^\nu] = 2ia^2
\textbf{I}^{\mu\nu}\eqno(19)$$

where $(\hat{\textbf{I}}^{\mu\nu})$ is the four-dimensional spin
operator. The space determined by the relations (17), (18a) and
(18b) will be called $\Gamma_4(\hat{ \textbf{x}})$- space.

    It follows from (19) that the eigenvalues of operators $\Delta
\hat{x}^0,
\Delta \hat{x}^1, \Delta \hat{x}^2, \Delta \hat{x}^3, \Delta
\hat{x}^4$ do not form the complete set. This set can be built out
of the eigenvalues of the interval operator $\Delta \hat{s}$ and
unit vector $textbf{N}$. By solving the equation for the
eigenfunctions $\phi_\lambda$ and eigenvalues of the operator
$\hat{\sigma}(N)=\frac{1}{a} \Delta \hat{s}(\textbf{N})$; that is

$$\hat{\sigma}(\textbf{N})\phi_\lambda = \lambda \phi_\lambda\eqno(20)$$

it is possible to find out the eigenvalues $\lambda$. They are
given as

$$\lambda =
\pm\sqrt{\textbf{N}^2}\hspace{.5em}for\hspace{.5em}\textbf{N}^2>0\eqno(21a)$$

$$\lambda =
\pm\sqrt{-\textbf{N}^2}\hspace{.5em}for\hspace{.5em}\textbf{N}^2<0\eqno(21b)$$

Thus, the eigenvalues of $\Delta \hat{s}$  are given by
$$\Delta s=\pm a\hspace{.5em}or\hspace{.5em}0$$

    It follows from (18a) (18b), (19) that the interval operators
$\Delta\hat{s}(\textbf{N}^\prime)$ and
$\Delta\hat{s}(\textbf{N}^{\prime\prime})$ for two nonparallel
directions $\textbf{N}^\prime$ and $\textbf{N}^{\prime\prime}$ do
not commute :
$$[\Delta\hat{s}(\textbf{N}^\prime),\Delta\hat{s}(\textbf{N}^{\prime\prime}]=a^2
\gamma^\mu \gamma^\nu (\textbf{N}^\prime\times
\textbf{N}^{\prime\prime})_{\mu\nu}\eqno(22)$$

where the symbol X represents the vector product. Hence, each
point of the quantized space $\Gamma_4(\hat{ \textbf{x}})$ can be
crossed only by one (though arbitrary) straight line.

    Regarding the ordering of events, Blokhintsev has chosen the sign for
the
interval in accordance with the concept of time $\tau$ and
distance $\ell$. For the time-like interval $\hat{s}=\hat{\tau},
\textbf{N}^2=1$, at each point, the rule
 $$\lambda =\pm 1, \hspace{.5em} \phi_\lambda \equiv \phi_\pm
 (\pm\textbf{N})\eqno(23)$$
gives two values of $\tau$, that is, $\tau = \pm a$, whereas for
the space-like interval $\hat{s} =\hat{\ell},\textbf{N}^2=-1,$
$$\lambda=+1,\hspace{.5em}\phi_\lambda =\phi_+
(\pm\textbf{N})\eqno(24)$$ only one sign is admitted, that is,
$\ell = a.$
    Thus, in accordance with this choice, at each point in
the space-like direction there can be only one ray$(\textbf{N})$,
while in time-like direction there can be two rays
$(\pm\textbf{N})$. Thereby the ordering of events is determined in
the space $\Gamma_4 (\hat{\textbf{x}})$. It is realized in the
same way as in the Minkowski space with the help of interval s and
unit vector $\textbf{N}$. The important difference is that only
one line (for $\textbf{N}^2=1$) and one ray (for
$\textbf{N}^2=-1$) are admitted at each point. The eigenvalue of
interval for $\textbf{N}^2=1$ coincides with time $\tau$ in the
reference frame, where $\textbf{N}$ = (1, 0, 0, 0) and for
$\textbf{N}^2=-1$ with length $\ell$ in the frame where
$\textbf{N}=(0,\vec{N})$. As to the interval $\Delta s= 0$,
$\textbf{N}^2=0$, it defines neither length nor time because at
$\Delta s= 0$ the operators $\hat{x}^0$ and $\hat{x}^h$(h=1,2,3)
do not commute with $\Delta\textbf{s}$ in any reference frame.
Therefore the seat of points separated by the light cone
$\textbf{N}^2=0$ is undermined.

    Similar quantized space-time has been considered by Namsrai (1985)
for the internal space-time $I_4$ of the space-time $\textbf{R}_4
= E_4 + I_4$ where $E_4$ is the external space-time. The
coordinates $\hat{x}^\mu \in \hat{R}_4$ ($\mu$ = 0, 1, 2, 3) are
given as
$$\hat{x}^\mu = x^\mu + r^\mu,\hspace{1em} x^\mu \in E_4
\hspace{1em}and \hspace{1em} r^\mu \in I_4$$

The quantization of space-time is realized in two possible ways,
namely,
$$Case\hspace{.5em} 1 :\hspace{1em} \hat{x}^\mu = x^\mu + \ell \gamma^\mu
\eqno(25a)$$
$$Case\hspace{.5em} 2 :\hspace{1em} \hat{x}^\mu = x^\mu + i\ell \gamma^\mu
\eqno(25b)$$ where $\ell$ is a fundamental length. Thus, $R_4$ is
quantized at small distances. For the first and second cases we
have, respectively,
$$[\hat{x}^\mu, \hat{x}^\nu] = 2i\ell^2
\sigma^{\mu\nu}\eqno(26a)$$
$$and\hspace{1em}[\hat{x}^\mu, \hat{x}^\nu] = \frac{2}{i}
\ell^2\sigma^{\mu\nu}\eqno(26b)$$
where\\
$$\sigma^{\mu\nu} = \frac{1}{2i} (\gamma^\mu \gamma^\nu -
\gamma^\nu \gamma^\mu)\eqno(27)$$

A mathematical procedure has been prescribed there to provide a
passage to the large scale from the small one. This procedure, the
averaging of coordinates $r^\mu=\ell\gamma^\mu$ or
$i\ell\gamma^\mu$ of the internal space$I_4$, is to trace the
$\gamma$ matrices. For example, in the case of first realization,
$$\langle\hat{x}^\mu\rangle_{\hat{R}_4} =
x^\mu,\hspace{.5em}\langle\hat{x}^\mu\hat{x}^\nu\rangle=x^\mu
x^\nu + 4\ell^2g^{\mu\nu}$$
$$\langle \hat{s^2}\rangle = \langle{\hat{x^0}}^{2} \rangle -
\langle{\hat{\vec{x}}}^{2} \rangle =s_0^2 +16\ell^2$$

where

$$s_0^2=x_0^2 - \vec{x}^2$$

For the second case,

$$\langle \hat{x}^\mu \hat{x}^\nu \rangle = x^\mu x^\nu - 4 \ell^2
g^{\mu\nu}$$
$$\langle \hat{s}^2 \rangle = s_0^2 - 16 \ell^2$$

It is to be noted that out of the two realizations we have to
choose only one realization for quantization of space-time. In
Blokhintsevs approach the quantization corresponds only to the
first case of Namsrais approach although there is an ambiguity in
the operator of interval $\Delta \hat{s}$ in the former one. In
fact, as it is evident from (18a) and (18b) that $\Delta \hat{s}$
differs for $\textbf{N}^2 = 1$ and $\textbf{N}^2 = 0$ from that
for $\textbf{N}^2 = -1.$

\section{MICRODOMAIN AS QUANTIZED FINSLER SPACE}
\noindent

    In De (1997) the microlocal space-time of extended hadrons has
been specified as a special Finsler space. The fundamental
function $F(\textbf{x},\mathbf{\nu})$ of this space is given as

$$\begin{array}{cc}
F^2(\textbf{x},\mathbf{\nu}) & =
\hat{g_{ij}}(\textbf{x},\mathbf{\nu})\nu^i \nu^j\\
where \hspace{.5em}\hat{g_{ij}}(\textbf{x},\mathbf{\nu}) & =
\eta_{ij}g(\textbf{x})\theta(\mathbf{\nu}^2)\\
\end{array}\Big\}\hspace{19.1em}\eqno(28)$$

Here, g is not, in general , the Finsler metric tensor, but simply
represents a homogeneous tensor of degree zero in $\mathbf{\nu}$,
which is used for the purpose of defining $\mathcal{F}$. The
Finsler metric can be obtained by using the following formula :

$$g_{ij}(\textbf{x},\mathbf{\nu})=\frac{1}{2} \frac{\partial^2
F^2(\textbf{x},\mathbf{\nu})}{\partial \nu^i
\partial \nu^j}\eqno(29)$$

The Finsler space introduced here is, in fact, in accord with
Riemanns original suggestion that the positive fourth root of a
fourth order differential form might serve as a metric function
(Riemann, 1854). That is, we can write

$$F(\textbf{x},\mathbf{\nu})=\{g_{\mu\nu\rho\sigma}(\textbf{x})\nu^\mu
\nu^\nu \nu^\rho \nu^\sigma\}^{1/4}\eqno(30)$$

In the present case the tensor field
$g=(g_{\mu\nu\rho\sigma}(\textbf{x}))$ is taken as

$$g_{\mu\nu\rho\sigma}(\textbf{x})=\{g(\textbf{x})\}^2
\eta_{\mu\nu} \eta_{\rho\sigma}\eqno(31)$$

Consequently,

$$F^4(\textbf{x},\mathbf{\nu})= \{g(\textbf{x})\}^2
\eta_{\mu\nu} \eta_{\rho\sigma}\nu^\mu \nu^\nu \nu^\rho
\nu^\sigma\} = \{g(\textbf{x})\mathbf{\nu}^2\}^2\eqno(32)$$

or,

$$ ds^4 = \{g(\textbf{x})\eta_{\mu\nu}dx^\mu dx^\nu\}^2\eqno(33)$$

>From this relation we have two possibilities, namely,

(i)the metric is given as

$$ds^2=g(\textbf{x})\eta_{\mu\nu}dx^\mu dx^\nu\eqno(34)$$

which corresponds to a Riemannian space with the metric tensor

$$g_{\mu\nu}(\textbf{x})= g(\textbf{x})\eta_{\mu\nu},$$

(ii)the metric

$$ds^2=g(\textbf{x})\theta(\mathbf{\nu}^2)\eta_{\mu\nu}dx^\mu
dx^\nu \eqno(35)$$

which corresponds to the Finsler space with the Fundamental
function (28). It is interesting to note that if we insist on the
condition of nonnegativeness of $ds^2$, then the second
possibility from (33) or (30) gives rise to a Finsler geometry and
we regard it as the geometry of the microlocal space-time of
hadronic matter-extension.

    Now, for quantization we write the interval of this Finsler space
in the following form :

$$ds_F = F(\textbf{x},d\textbf{x})=L_\mu
dx^\mu\eqno(36)$$

where $L_\mu$ is a zero-order form in $\nu^\mu\equiv
\frac{dx^\mu}{ds_F}$ . In fact,$L_\mu$  and $\L^\mu$ are given by
(for nonnull $ds_F$)

$$L^\mu =
g(\textbf{x})\theta(\mathbf{\nu}^2)\nu^\mu,\hspace{1em}and\hspace{1em}L_\mu
=g(\textbf{x})\theta(\mathbf{\nu}^2)\eta_{\mu\nu}\nu^\nu\eqno(37)$$

For null $ds_F$ we can take

$$L^\mu = k g(\textbf{x})\theta(\mathbf{\nu}^2)
\frac{dx^\mu}{dt},\hspace{1em}and\hspace{1em}L_\mu =k
g(\textbf{x})\theta(\mathbf{\nu}^2)\eta_{\mu\nu}\frac{dx^\nu}{dt}\eqno(38)$$

where t is a parameter and k is (-1) p-homogeneous form in the
differential$dx^\mu$ (that is, a positively homogeneous function
of degree -1). k is also supposed to be positive definite for all
directions. Then, for nonnull $ds_F$ we have

$$\textbf{L}^2 =\eta_{\mu\nu}L^\mu
L^\nu=g(\textbf{x})\theta(\mathbf{\nu}^2)F(\textbf{x},\mathbf{\nu})=g(\textbf{x})\theta(\mathbf{\nu}^2)\eqno(39)$$

(using (28) and noting
that$F(\textbf{x},\mathbf{\nu})=F(\textbf{x},\frac{d\textbf{x}}{ds_F})=1$
For null $ds_F$ ,

$$\textbf{L}^2 =\eta_{\mu\nu}L^\mu L^\nu= k^2
g^2(\textbf{x})\theta^2(\mathbf{\nu}^2)\eta_{\mu\nu}\frac{dx^\mu}{dt}\frac{dx^\nu}{dt}=0\eqno(40)$$

This property of $L^\mu$ expresses the Finslerian character of the
space-time given by the interval $ds_F$ in (36).

    The quantum generalization of this Finsler space is made through
the change of coordinate differentials in (36) by the finite
operators

$$\Delta \hat{x}^\mu =\epsilon \sqrt{\theta(\mathbf{\nu}^2)}
\gamma^\mu(\textbf{x})\eqno(41)$$

where $\gamma^\mu(\textbf{x})\hspace{.5em}(\mu =0,1,2,3)$ are
Dirac matrices for the associated curved space(Riemannian) of the
finsler space. In fact, the present Finsler space$ F_4 = (M_4, F)$
that describes the microlocal space-time of extended hadrons is a
simple type of$(\alpha, \beta)$ - metric Finsler space. The
fundamental function F of this space can be written as
$$F(\textbf{x},\mathbf{\nu}) = \alpha
\sqrt{\theta(\mathbf{\alpha}^2)} = \alpha \beta^0
\sqrt{\theta(\mathbf{\alpha}^2)}$$

where
$$\alpha(\textbf{x},\mathbf{\nu})=\sqrt{g(\textbf{x})\eta_{ij} \nu^i \nu^j
},\hspace{.5em} g(\textbf{x}) > 0,$$

and $\beta(\textbf{x},\mathbf{\nu})$ is a differential 1-form.
Obviously, the function F is (1) p-homogeneous in $\alpha$ and
$\beta$. The associated space $R_4 = (M_4,\alpha)$ is Riemannian.
This space is conformal to the Minkowski space-time. The matrices
$\gamma^\mu(\textbf{x})$ are related to the flat space (Dirac)
matrices through the vierbein $V_\alpha^\mu(\textbf{x})$ as
follows :

$$\gamma^\mu(\textbf{x}) =
V_\alpha^\mu(\textbf{x})\gamma^\alpha\eqno(42)$$

For the present case, the vierbein fields
$V_\alpha^\mu(\textbf{x})$ and $V_\mu^\alpha(\textbf{x})$ are
given as

$$V_\alpha^\mu(\textbf{x}) = \{{g(\textbf{x})}\}^{-1/2}
\delta_\alpha^\mu,\hspace{1em}and\hspace{.5em}V_\mu^\alpha(\textbf{x})
=\{{g(\textbf{x})}\}^{1/2}\delta_\mu^\alpha\eqno(43)$$

With the operators (41) we have the operator of interval as

$$\Delta \hat{s}_F = L_\mu \Delta \hat{x}^\mu\eqno(44)$$

Also, from (41),(42), and (43) it follows that

$$[\Delta \hat{x}^\mu, \Delta \hat{x}^\nu] = 2i\epsilon^2
\frac{\theta{(\mathbf{\nu}^2})}{g(\textbf{x})}
\sigma^{\mu\nu}\eqno(45)$$

Defining the operator $\hat{\sigma}(\textbf{L}) = \frac{\Delta
\hat{s}_F}{\epsilon}$, we have

$$\hat{\sigma}(\textbf{L}) =\sqrt{\theta(\mathbf{\nu})^2} L_\mu
\gamma^\mu(\textbf{x})=\sqrt{\frac{\theta({\mathbf{\nu}^2})}{g(\textbf{x})}}
L_a \gamma^a\eqno(46)$$

where $\gamma^A$ are flat space Dirac matrices. Also, it follows
that

$$\hat{\sigma^2} = \frac{\theta({\mathbf{\nu}^2})}{g(\textbf{x})}
\textbf{L}^2 = 1\hspace{1em}for\hspace{1em}\textbf{L}^2 \neq
0\eqno(47a)$$

$$\hat{\sigma^2} = \frac{\theta({\mathbf{\nu}^2})}{g(\textbf{x})}
\textbf{L}^2 = 0\hspace{1em}for\hspace{1em}\textbf{L}^2 =
0\eqno(47b)$$

Thus, the eigenvalues of $\hat{\sigma}$ are $\pm 1
\hspace{.5em}and\hspace{.5em} 0$. Consequently, those of the
interval operator $\Delta \hat{s}_F$ are $\pm \epsilon$ and $0$.
The ordering of events in this space can be determined as in the
case of Minkowski (Finsler) space discussed above.

\section{FIELD EQUATION IN FINSLER SPACE}
\noindent

    We shall now derive quantum field (or wave) equation in the
Finsler space of microdomain through its quantum generalization.
The quantization is admitted in two steps. From the property of
the field function as given in (2) and (11) one can write

$$ \delta\psi = dx^\mu \partial_\mu\psi(\textbf{x},\mathbf{\nu})+ dv^\ell
\partial_\ell^\prime  =
-\frac{i\epsilon m c}{\hslash}\sqrt{\theta(\mathbf{\nu})^2}
\psi(\textbf{x},\mathbf{\nu})\eqno(48)$$

 As the first step, the differentials $dx^\mu$ are quantized to $\Delta
\hat{x}^\mu$ as
given in (41). Conseqently, the classical field function
$\psi(\textbf{x},\mathbf{\nu})$ transforms into a spinor
$\psi_{\alpha} (\textbf{x},\mathbf{\nu})$ whose equation becomes

$$\epsilon \sqrt{\theta(\mathbf{\nu})^2}
\gamma_{\alpha\beta}^\mu(\textbf{x})\partial_\mu
\psi_\beta(\textbf{x},\mathbf{\nu}) + d\nu^\ell
\partial_\ell^\prime \psi_\alpha (\textbf{x},\mathbf{\nu}) = -
\frac{i\epsilon m c}{\hslash}\sqrt{\theta(\mathbf{\nu})^2}
\psi_\alpha(\textbf{x},\mathbf{\nu})\eqno(49)$$

In the second step, the differentials dnl are quantized by noting
first that since the neighbouring points $
(\textbf{x},\mathbf{\nu})$ and $(\textbf{x} +
d\textbf{x},\mathbf{\nu} + d\mathbf{\nu})$ lie on the autoparallel
curve of the Finsler space the relation (1) between the
differentials $d\nu^\ell$ and $dx^\mu$ must hold.  Therefore, the
quantized differentials $\Delta \hat{\nu}^\ell$ should be given as

$$\Delta \hat{\nu}^\ell = - \epsilon \sqrt{\theta(\mathbf{\nu}^2)}
 \gamma_{h\mu}^\ell (\textbf{x},\mathbf{\nu}) \nu^h
 \gamma^\mu(\textbf{x})\eqno(50)$$
and consequently, the field becomes a bispinor (this
nomenclature is only formal and is for convenience)
$\psi_{\alpha\beta}(\textbf{x},\mathbf{\nu})$. Then the resulting
equation for $\psi_{\alpha\beta}$ in the Finsler space becomes

$$i\hbar\{\gamma_{\alpha\beta^\prime}^\mu(\textbf{x})\partial_\mu
\psi_{\beta^\prime\beta} (\textbf{x},\mathbf{\nu}) -
\gamma_{\beta\beta^\prime}^\mu (\textbf{x}\gamma_{h\mu}^\ell
(\textbf{x},\mathbf{\nu}) \nu^h \partial_\ell^\prime
\psi_{\alpha\beta^\prime} (\textbf{x},\mathbf{\nu}) = mc
\psi_{\alpha\beta} (\textbf{x},\mathbf{\nu})\}\eqno(51a)$$

or, in compact form,

$$i\hslash\gamma^\mu(\textbf{x})(\partial_\mu - \gamma_{h\mu}^\ell
(\textbf{x},\mathbf{\nu}) \nu^h \partial_\ell^\prime)
\psi(\textbf{x},\mathbf{\nu}) = mc
\psi(\textbf{x},\mathbf{\nu})\eqno(51b)$$

where it is to be remembered that the first and the second
operators on the L.H.S. operate on the first and the second spinor
indices of the bispinor $\psi(\textbf{x},\mathbf{\nu}) =
\psi_{\alpha\beta}(\textbf{x},\mathbf{\nu})$ respectively.

\section{DIRAC EQUATION IN ROBERTSON-WALKER SPACE-TIME}
\noindent

    We shall now find the field equation (for lepton) in the
associated Riemannian space of the Finsler space from the above
equation of the bispinor field. For this purpose one can first
specify the function $g(\textbf{x})$ in the metric (28) as

$$ g(\textbf{x}\equiv F(t) = exp(\pm b_0
x^0)\hspace{0.5em}or\hspace{0.5em}(b_0
x^0)^n\hspace{0.5em}or\hspace{0.5em}(1 + b_0 x^0)^n\eqno(52)$$

where$\hspace{0.5em}x^0 = ct$

In this case, the connection coefficients $\gamma_{h\mu}^\ell
(\textbf{x},\mathbf{\nu})$ are separated as

$$\gamma_{h\mu}^\ell (\textbf{x},\mathbf{\nu}) =
\zeta(t)\gamma_{h\mu}^\ell(\mathbf{\nu})\eqno(53)$$

where
$$2b_0 c \zeta(t) = F^\prime/F(t)\eqno(54)$$

In fact, for the present Finsler space it can be easily seen that
$\gamma_{h\mu}^\ell$ are independent of the directional arguments.
Also, if we calculate $G_{hk}^\ell =\frac{\partial^2
G^\ell}{\partial\nu^h\partial\nu^k}$ with $G^\ell = \frac{1}{2}
\gamma_{ij}^\ell(\textbf{x},\mathbf{\nu})\nu^i\nu^j$ it will be
found that $G_{hk}^\ell$ are independent of $\mathbf{\nu}$. Such
types of Finsler spaces are called affinely - connected spaces or
Berwald spaces (Rund,1959).

    Now, the field equation (51a) can be written in the following form
:
$$ i\hslash \{\gamma^\mu (\textbf{x}) \partial_\mu
\psi(\textbf{x},\mathbf{\nu}) - \zeta(t)\gamma_{h\mu}^\ell
\nu^h(\psi(\textbf{x},\mathbf{\nu})\overleftarrow{\partial_\ell^\prime}\gamma^{\mu^T}(\textbf{x}))\}
= m c \psi(\textbf{x},\mathbf{\nu})\eqno(55)$$

where the bispinor$\psi(\textbf{x},\mathbf{\nu})$ is represented
here as a 4 x 4 matrix. The vierbein fields $V_a^\mu$ and the
inverse vierbein fields $V_\mu^a$ satisfy

$$V_a^\mu(\textbf{x})V_\mu^b(\textbf{x}) = \partial_a^b\eqno(56)$$

For the present case, these fields are diagonal and are given by

$$V_a^\mu(\textbf{x}) =e(t) \partial_a^\mu, \hspace{1em}
V_\mu^a(\textbf{x}) = \frac{1}{e(t)} \partial_\mu^a\eqno(57)$$

where $e(t)=\{F(t)\}^{1/2}$
Then, the equation (55) becomes

$$i\hslash e(t)\{\gamma^\mu (\textbf{x}) \partial_\mu
\psi(\textbf{x},\mathbf{\nu}) -
\zeta(t)\psi(\textbf{x},\mathbf{\nu})\overleftarrow{\partial_\ell^\prime}\gamma^{\mu^T}\gamma_{h\mu}^\ell
\nu^h\}= m c \psi(\textbf{x},\mathbf{\nu})\eqno(58)$$

where $\gamma^\mu$ are now the flat space Dirac matrices.

    Let us now decompose $\psi(\textbf{x},\mathbf{\nu})$ in the following
way (De, 1997) :

$$\psi(\textbf{x},\mathbf{\nu}) = \psi_1(\textbf{x}) \times
\phi^T(\mathbf{\nu}) + \psi_2(\textbf{x}) \times
\phi^{c^T}(\mathbf{\nu})\eqno(59)$$

where the spinors $\psi_1(\textbf{x})$ and $\psi_2(\textbf{x})$
are eigenstates of $\gamma^0$ with eigenvalues +1 and -1
respectively. Also, $\phi(\mathbf{\nu})$ and
$\phi^c(\mathbf{\nu})$ satisfy, respectively, the following
equations.
$$i\hslash\gamma^\mu \gamma_{h\mu}^\ell \nu^h \partial_\ell^\prime
\phi(\mathbf{\nu}) =(Mc - \frac{3i\hslash
b_0}{2})\phi(\mathbf{\nu})\eqno(60a)$$

$$i\hslash\gamma^\mu \gamma_{h\mu}^\ell \nu^h \partial_\ell^\prime
\phi(\mathbf{\nu}) =(Mc + \frac{3i\hslash
b_0}{2})\phi^c(\mathbf{\nu})\eqno(60b)$$

Then, it is easily seen that the field
$\psi(\textbf{x},\mathbf{\nu})$ satisfies the Dirac equation in
the "$\textbf{x}$-space", the associated Riemannian space, which
is, in fact, a space-time conformal to the Minkowski flat space.
Here, M appears as a constant in the process of separation of the
equation (58) and this can be considered as a manifestation of the
anisotropic Finslerian character of the microdomain. The equation
for   $\psi(\textbf{x},\mathbf{\nu})$ is

$$i\hslash (\gamma^\mu \partial_\mu + \frac{3b_0}{2}
\zeta(t)\gamma^0)\psi(\textbf{x},\mathbf{\nu}) =
\frac{c}{e(t)}(m+M\zeta(t)e(t))\psi(\textbf{x},\mathbf{\nu})\eqno(61)$$

Consequently, the Dirac equation for the Robertson-Walker (RW)
space-time can be obtained by a pure-time transformation given as

$$\frac{dt}{e(t)}=\sqrt{F(t)}dt
=dT\hspace{0.5em}or,\hspace{0.5em}\int\frac{dt}{e(t)}=
T\eqno(62)$$ where T is the cosmological time. With the scale
factor $R(T)\equiv \sqrt{F(t)}$, this transformation can also be
written as

$$t = \int \frac{dT}{R(T)}\eqno(63)$$

Here, in the Dirac equation (61) (and also in that for RW
space-time) an additional mass term (if$ M \neq 0$) appears and it
is time-dependent. In fact, the mass of the particle is$ m +
M\zeta(t) e(t)$ where m is the "inherent" mass of the particle as
stated earlier. The time-dependent part of the mass, expressed in
terms of cosmological time is found to be dominant in the very
early universe and has significant effect in that era (De, 1993).
It is pointed out  here that the "$\mathbf{\nu}$-part" in the
decomposition of$\psi(\textbf{x},\mathbf{\nu})$ can give rise to
an additional quantum number if it represents the field of the
constituent-particle in the hadron configuration (De, 1997). On
the other hand, for other cases, time-dependence of the masses is
the only physical consequence that is manifested by the
$\mathbf{\nu}$-variable-dependence of the bispinor field. The
usual field for the $(\textbf{x})$-space can be obtained by an
"averaging procedure" such as

$$\psi(\textbf{x}) = \int
\psi(\textbf{x},\mathbf{\nu})\chi(\mathbf{\nu})d^4\nu\eqno(64)$$

where $\chi(\mathbf{\nu})$ is a (spinor) probability density or a
weight function. The field $\psi(\textbf{x})$ clearly satisfies
the Dirac equation (61) in the $(\textbf{x})$-space.

    The Dirac equation for the local inertial frame (the Minkowski
flat space) can be recovered from (61) by using the vierbein

$$ V_\mu^\alpha(\textbf{X}) = (\frac{\partial y_x^\alpha}{\partial
x^\mu})_{\textbf{x}=\textbf{X}}\hspace{0.5em} (\alpha =
0,1,2,3)\eqno(65)$$

which connect the curved space-time with the flat one in normal
coordinates $y_x^\alpha$ (with index $\alpha$ referring to the
local inertial frame) at the point $\textbf{X}$. In the present
case, the index $\mu$ is associated with the conformal Minkowski
space-time which becomes the RW space-time by a pure-time
transformation. For fixed $y_x^\alpha$, the effect of changing
$x^\mu$ is given by

$${V_\mu^\alpha} \rightarrow \frac{\partial x^{\nu}}{\partial x^{\prime
\mu}}
V_\nu^\alpha \hspace{0.5em}\eqno(66)$$

On the other hand, $y_x^\alpha$ may be changed by Lorentz
transformation $\Lambda_\beta^\alpha (\textbf{X})$, and in this
case the vierbeins $V_\mu^\alpha(\textbf{X})$ are changed
to$\Lambda_\beta^\alpha (\textbf{X})V_\mu^\beta(\textbf{X})$
keeping the metric of the curved space-time invariant.

    With vierbeins (57) for the present case, one can obtain from (61)
the Dirac equation in local inertial frame in normal coordinate
system as

$$i\hslash \gamma^\alpha \partial_\alpha
\psi(\textbf{x},\mathbf{\nu})= mc
\psi(\textbf{x},\mathbf{\nu})\eqno(67)$$

if one neglects the extremely small second terms both from the
left and right sides of (61). Of course, the mass term $M \zeta(t)
e(t)$  may be retained here, and although it has no appreciable
effect on the mass of the particle in the present era its
dominance in the very early epoch is very significant (De, 1993,
1999).

    Actually, in laboratory space-time (Minkowskian) one can decompose
the field as

$$\psi(\textbf{x},\mathbf{\nu}) = \psi(\textbf{x}) X
\phi^T(\mathbf{\nu})=
\psi(\textbf{x})\phi^T(\mathbf{\nu})\eqno(68)$$

where $\phi(\mathbf{\nu})$  satisfies the equation

$$i\hslash\gamma^\mu \gamma_{h\mu}^\ell(\mathbf{\nu})\nu^h
\partial_\ell^\prime
\phi(\mathbf{\nu}) =Mc \phi(\mathbf{\nu})\eqno(69)$$

The decomposition (68) corresponds to that in (59) for the case
$\phi(\mathbf{\nu}) = \phi^c(\mathbf{\nu})$ and for flat
space-time $(b_0 \rightarrow 0)$. With this decomposition, it is
easy to see that (from equation (58))$\psi(\textbf{x})$ satisfies
the Dirac equation in flat space-time :

$$i\hslash\gamma^\mu \partial_\mu \psi(\textbf{x}) = mc
\psi(\textbf{x})\eqno(70)$$

Here, of course,$b_0 \rightarrow 0$ (and consequently
$F(t)\rightarrow 1)$ makes $M \rightarrow 0$, as we shall later
see that M is proportional to $b_0$. The bispinor
$\psi(\textbf{x},\mathbf{\nu})$ which satisfies the Dirac equation
(67) in the local inertial frame (the Minkoski space-time) reduces
to a spinor through the  averaging procedure (64) and this spinor
satisfies the Dirac equation in local frame with or without the
time-dependent mass term. Thus, the time-dependence of particle
masses is the manifestation of the Finslerian character of the
space-time whose underlying manifold (the $\psi(\textbf{x})$-space
or the associated Riemannian space) is a curved one. In the
present case, this space is a RW space-time (obtained through
pure-time transformation) which is the background space-time of
our universe.

\section{HOMOGENEOUS SOLUTIONS FOR $\phi(\mathbf{\nu})$ AND
$\phi^c(\mathbf{\nu})$}
\noindent
    We seek a class of solutions for $\phi(\mathbf{\nu})$ and
$\phi^c(\mathbf{\nu})$  which are
homogeneous of degree zero. In fact, the metric tensors of the
Finsler space and the fundamental function are homogeneous
functions of degree zero and one, respectively, in the directional
arguments. Therefore, one can argue that only this class of
homogeneous solutions is physically relevant. For the Finsler
space that we are considering, the equations for
$\phi(\mathbf{\nu})$and $\phi^c(\mathbf{\nu})$ for such type of
solutions become (from eqn.(60))

$$i\hslash b_0 \sum_{\ell = 1}^3 \gamma^\ell(\nu^\ell
\frac{\partial}{\partial
\nu^0} + \nu^0\frac{\partial}{\partial
\nu^\ell})\phi(\mathbf{\nu}) = -(Mc - \frac{3i\hslash
b_0}{2})\phi(\mathbf{\nu})\eqno(71a)$$ \hspace{1em}and\hspace{1em}

$$i\hslash b_0 \sum_{\ell = 1}^3 \gamma^\ell (\nu^\ell
\frac{\partial}{\partial
\nu^0} + \nu^0\frac{\partial}{\partial
\nu^\ell})\phi^c(\mathbf{\nu}) = -(Mc + \frac{3i\hslash
b_0}{2})\phi^c(\mathbf{\nu})\eqno(71b)$$

The general form of the solutions of these equations, which are
homogeneous of degree zero in n is given as
$$\phi(\mathbf{\nu}) =
\{f_1 (\frac{\vec{\nu^2}}{\nu^{0^2}}) + \frac{i\vec{\gamma}
.\vec{\nu}}{\nu^0}f_2
(\frac{\vec{\nu^2}}{\nu^{0^2}})\}\omega^b\eqno(72)$$

where $\omega^b$ is a four-component (arbitrary) spinor
independent of $\mathbf{\nu}$ Here, $f_1$ and $f_2$ are functions
of the homogeneous variable (of degree zero in $\mathbf{\nu}$)
$x=\vec{\nu^2}/\nu^{0^2}.$ These functions satisfy the following
coupled equations

$$2f_1^\prime(x)(1-x) =ikf_2(x)\eqno(73a)$$

$$(x-3)f_2(x)+2x(x-1)f_2^\prime(x) = -ikf_1(x)\eqno(73b)$$

where k is a real or complex constant. In fact, it is given by

$$i\hslash b_0 k =-Mc + \frac{3i\hslash b_0}{2}\eqno(74)$$

For $\phi^c(\mathbf{\nu})$ whose form of solution is also given by
(72), the constant k  is to be replaced by $k^\prime$, and it is
given by

$$i\hslash b_0 k^\prime = -Mc - \frac{3i\hslash b_0}{2}\eqno(75)$$

Thus, we have  $$Re \{k\} =3/2, \hspace{1em}Re\{k^\prime\}=
-3/2,\hspace{1em}Im\{k\}=Im\{k^\prime\}=Mc/\hslash b_0\eqno(76)$$

For M = 0, k and $k^\prime$ are real. For the equation (69), the
solution for $\phi(\mathbf{\nu})$ is also of the form (72) but in
this case k is purely imaginary, and it is given by $k
=iMc/\hslash b_0.$

    Now, it is easy to see that $f_1(x)$ and $f_2(x)$ individually satisfy
the following second order differential equations

$$4(x-1)^2 \{x f_1^{\prime\prime}(x) + (3/2) f_1^\prime(x)\} + k^2
f_1(x) =0\eqno(77a)$$

$$4(x-1)^2 \{x f_2^{\prime\prime}(x) + (5/2) f_2^\prime(x)\}
+\{k^2-2(1-x)\}f_2(x) = 0\eqno(76b)$$

With the substitution $y = x^{-1/2}$, we get the equation for
$f_1(x)$ from (77a) as

$$\frac{d^2f_1}{dy^2} + \frac{k^2 f_1}{(1-y^2)^2} =
0\hspace{1em}(y^2\neq 1)\eqno(78)$$

Again, by the substitution

$$f_1 = (1-y^2)^{1/2} \nu\hspace{1em}(y^2\neq 1)\eqno(79)$$

we have the equation for $\nu$ as

$$ (1-y^2)\frac{d^2\nu}{dy^2} - 2y\frac{d\nu}{dy} +
\frac{k^2-1}{1-y^2}\nu = 0\eqno(80)$$

The solutions for $\nu$ can be identified for $y^2 > 1$ and $y^2 <
1.$ In fact, for $y^2 > 1$, one can make the substitution

$$f_1 = (y^2-1)^{1/2} \nu\eqno(81)$$

in (78) to arrive at the same-equation (80). For $y^2 > 1$, the
solution for $\nu$ is the associated Legendre function of second
kind $Q_0^\mu(y)$ where $\mu = (1 - k^2)^{1/2}.$ It is to be noted
that for $M = 0$, we have $k^2 = k^{\prime^2} = 9/4$ and
consequently $\mu = i\sqrt{5}/2$.

    Now, by using the following integral representation of $Q_0^\mu(y)$
(Erdlyi,1953)

$$Q_0^\mu(y) = (1/2) e^{\pi\mu i}\Gamma(\mu + 1)(y^2-1)^{-\mu/2}
\int_0^\pi (y+cos\hspace{0.2em} t)^{\mu-1} sin\hspace{0.2em}t
dt,\hspace{1em} for\hspace{1em}Re(\mu+1) > 0$$

and with the substitution $y + cos\hspace{0.2em} t = \xi$ we can
have the solution for $\nu$ as

$$\nu = Q_0^\mu(y) = (1/2\mu)e^{\pi\mu i}\Gamma(\mu +
1)(y^2-1)^{-\mu/2}\{(y +1)^\mu - (y-1)^\mu \}$$

Thus, apart from the inessential constant one can take the
solution for $f_1$ for $y^2 > 1$ (that is, for $y > 1$ or $y <
-1$) as

$$ f_1 = \frac{(y^2-1)^{1/2}}{2\mu} \{ (\frac{y+1}{y-1})^{\mu/2} -
(\frac{y-1}{y+1})^{\mu/2}\}\eqno(82)$$ For $y^2 < 1$, (that is,
for$ - 1 < y < 1$), it is easy to see that

$$ f_1 = \frac{(1-y^2)^{1/2}}{2\mu} \{ (\frac{1+y}{1-y})^{\mu/2} -
(\frac{1-y}{1+y})^{\mu/2}\}\eqno(83)$$ In alternative forms we can
write

$$f_1 =\frac{\sqrt{y^2-1}}{d}
sin\{(d/2)ln(\frac{y+1}{y-1})\}\hspace{1em}for \hspace{0.5em} y >
1 \hspace{0.5em}or\hspace{0.5em}y < -1\eqno(84a)$$

$$f_1 =\frac{\sqrt{1-y^2}}{d}
sin\{(d/2)ln(\frac{1+y}{1-y})\}\hspace{1em}for \hspace{0.5em}-1 <
y < 1\eqno(84b)$$

where $d =\sqrt{5}/2$

Now, $f_2$ can be found from (73) and (84), and it is found to be

$$ f_2 =\frac{y\sqrt{y^2-1}}{ik}[cos\{(d/2)ln(\frac{y+1}{y-1})\}
- (y/d)sin\{(d/2)ln(\frac{y+1}{y-1})\}]\hspace{1em}for
\hspace{.5em} y > 1 \hspace{.5em}or\hspace{0.5em}y < -1
\eqno(85a)$$

$$f_2 =\frac{y\sqrt{1-y^2}}{ik}[cos\{(d/2)ln(\frac{1+y}{1-y})\}
- (y/d)sin\{(d/2)ln(\frac{1+y}{1-y})\}]\hspace{1em}for
\hspace{0.5em}-1 < y <1\eqno(85b)$$

Here,   y is given by  $$y  =
\nu^0/\sqrt{\vec{\nu}^2}\eqno(86)$$which is a homogeneous function
of degree zero in the directional variable $\nu$. For the case of
$M = 0$, the solution for $\phi(\mathbf{\nu})$ from (72) can be
written as

$$\phi(\mathbf{\nu}) = F(\nu^0,\vec{\nu}) \omega
^b\eqno(87)$$

where

$$F(\nu^0,\vec{\nu}) = \sqrt{\frac{\nu^{0^2}}{\vec{\nu}^2}-1}[(1/d)
sin\{(d/2) ln
\frac{\nu^0+\sqrt{\vec{\nu}^2}}{\nu^0-\sqrt{\vec{\nu}^2}}\} +
\frac{\vec{\gamma}.\vec{\nu}}{k\sqrt{\vec{\nu}^2}}\{cos( (d/2)ln
\frac{\nu^0+\sqrt{\vec{\nu}^2}}{\nu^0-\sqrt{\vec{\nu}^2}}) -$$
$$\frac{\nu^0}{d\sqrt{\vec{\nu}^2}}sin((d/2) ln
\frac{\nu^0+\sqrt{\vec{\nu}^2}}{\nu^0-\sqrt{\vec{\nu}^2}})\}]\hspace{0.5em}\nu^0
> \sqrt{\vec {\nu^2}} \hspace{0.5em}or\hspace{0.5em}\nu^0 <
- \sqrt{\vec {\nu^2}}\eqno(88a)$$
and

$$F(\nu^0,\vec{\nu}) = \sqrt{1-\frac{\nu^{0^2}}{\vec{\nu}^2}}[ (1/d)
sin\{(d/2) ln \frac{\sqrt{\vec{\nu}^2}+\nu^0}{\nu^0-\sqrt{\vec{\nu}^2}}\}
+
\frac{\vec{\gamma}.\vec{\nu}}{k\sqrt{\vec{\nu}^2}}\{cos( (d/2)ln
\frac{\sqrt{\vec{\nu}^2}+\nu^0}{\sqrt{\vec{\nu}^2}-\nu^0}) -$$
$$\frac{\nu^0}{d\sqrt{\vec{\nu}^2}}sin((d/2) ln
\frac{\sqrt{\vec{\nu}^2}+\nu^0}{\sqrt{\vec{\nu}^2}
-\nu^0})\}]\hspace{0.5em} - \sqrt{\vec {\nu^2}} < \nu^0 <
\sqrt{\vec {\nu^2}}\eqno(88b)$$

The solution for $\phi^c(\mathbf{\nu})$ can be obtained from (88a)
and (88b) by changing $k = 3/2$ to  $k^\prime = - 3/2.$ This can
also be achieved in the solution by changing $\vec\nu$ to
$-\vec\nu$ (keeping the constant k unchanged).  That is, the
solution for $\phi^c(\mathbf{\nu})$ is actually given by

$$\phi^c(\mathbf{\nu}) = F(\nu^0, -\vec\nu)\omega_c^b\eqno(89)$$

where $\omega_c^b$ is an arbitrary four-component spinor
independent of $\mathbf{\nu}$ and may be chosen as

$$\omega_c^b = i\gamma^2(\omega^b)^\ast\eqno(90)$$

It is to be noted that the solutions for $\phi(\mathbf{\nu})$ and
$\phi^c(\mathbf{\nu})$ at $y^2 = 1$, that is, at $\nu^0= \pm
\sqrt{\vec\nu^2}$ can be obtained from (87), (88a) and (88b) by
taking the left and right limits at those points. Obviously, these
limits are zero, and consequently,$\phi(\mathbf{\nu})$ and
$\phi^c(\mathbf{\nu})$ tend to zero at $\nu^0= \pm
\sqrt{\vec\nu^2}.$ This results is also in consistent with the
relations. It is easy to see that $\phi(\mathbf{\nu})$ (and also
$\phi^c(\mathbf{\nu})$) is finite as $\vec\nu^2 \rightarrow 0$
(that is, $y \rightarrow \infty)$. In fact
$\phi(\mathbf{\nu})\rightarrow \omega^b$ as $\vec\nu^2 \rightarrow
0$, that is, as $\mathbf{\nu} \rightarrow  (\nu^0,0,0,0).$ Also,
as $\mathbf{\nu} \rightarrow (0,\vec \nu)$, that is, as $y
\rightarrow 0$, $\phi(\mathbf{\nu})\rightarrow
\frac{\vec{\gamma}.\vec{\nu}}{k\sqrt{\vec{\nu}^2}} \omega^b$ and
$\phi^c(\mathbf{\nu})\rightarrow -
\frac{\vec{\gamma}.\vec{\nu}}{k\sqrt{\vec{\nu}^2}} \omega_c^b.$

The following relation also holds good :

$$\phi^c(\mathbf{\nu}) = i \gamma^2 \phi^\ast
(\mathbf{\nu})\eqno(91)$$

with the following representation of $\gamma$-matrices

$$\gamma^0 = \left(%
\begin{array}{cc}
  1 & 0 \\
  0 & -1 \\
\end{array}%
\right)\hspace{1em}and \hspace{1em}\vec\gamma =\left(%
\begin{array}{cc}
  0 & \vec\sigma \\
  -\vec\sigma & 0 \\
\end{array}%
\right)\eqno(92)$$

$\vec\sigma = (\sigma^1,\sigma^2,\sigma^3)$ being Paulis spin
matrices.
    For $M \neq 0$, that is, for $Im\{k\} = Im\{k^\prime\} \neq 0$, the
solution
of (80) is also given by $\nu = Q_0^\mu (y)$ but in this case
$\mu$ is a complex number. It is given by

$$\mu = \sqrt{1-\{(3/2) + id\}^2} \eqno(93)$$

where $Im\{k\} = Im\{k^\prime\} = d$

>From (76) we see that the real parts of k and $k^\prime$ are fixed
but their imaginary parts are arbitrary. One can choose these
imaginary parts to be of the same order with their real parts.
Presently, we take

$$Im\{k\} = Im\{k^\prime\} = d = 3/\sqrt{13} \eqno(94)$$

and therefore       $$\mu = 1 - \frac{9i}{2\sqrt{13}}\eqno(95)$$

With this changed $\mu$, the solutions for $f_1$ and $f_2$ are as
above. It is to be noted that this choice of the value for d can
give rise to the finite limits for $f_1$ and $f_2$ as $y
\rightarrow \infty$, and also for $y \rightarrow 0.$ The functions
$f_1$ and $f_2$ at $y = \pm 1$ are given, as before, by their
limits as $y \rightarrow \pm 1$ from left as well as from right.
It is easy to see that these limits are also finite for this value
of d.

    With this value of d, we can obtain the following relation from
(76) :

$$\frac{3\hslash b_0}{\sqrt{13}} = Mc \hspace{1em}or,
\hspace{1em}(0.832)\hslash
b_0= Mc\eqno(96)$$

This relation connects the mass parameter M with the parameter
$b_0$ of the space-time. Although, the terms containing $b_0$ and
M may be neglected to obtain the usual field equations in local
inertial frame of the "$\textbf{x}$-space" (the flat space) but it
is seen that the $\nu$-dependent spinor $\phi(\mathbf{\nu})$ (for
the case $M = 0$) can give rise to an additional quantum number
which can generate the internal symmetry of hadrons (De, 1997).

\section{EXTERNAL ELECTROMAGNETIC FIELD}
\noindent

    The field equation for a particle in an external electromagnetic
field can be deduced from a similar property of the fields on the
autoparalled curve of the Finsler space as for case of a free
field considered above. This property is now expressed by the
following relation :

$$\delta (\psi\chi) = \frac{ie}{\hslash c} \Bigl\{A_\mu
(\textbf{x}, \mathbf{\nu})dx^\mu \Bigl \}\psi\chi = -
\frac{i\epsilon mc}{\hslash} \sqrt{\theta (\mathbf{\nu}^2)}
\psi\chi \eqno(97)$$

where $\chi(\textbf{x}, \mathbf{\nu})$ is a scalar function which
may be regarded as a phase factor for the  field $\psi
(\textbf{x}, \mathbf{\nu})$. The vector field  $A_\mu(\textbf{x},
\mathbf{\nu})$ represents the external electromagnetic field. From
(97), we have

$$dx^\mu\Bigl\{ \partial_\mu + \frac{ie}{\hslash c} A_\mu (\textbf{x},
\mathbf{\nu})+ \frac{1}{\chi} \partial_\mu \chi \Bigl \} \psi +
d\nu^\ell \Bigl\{\partial_\ell^\prime +
\frac{1}{\chi}\partial_\ell^\prime \chi \Bigl \}\psi = -
\frac{i\epsilon mc}{\hslash} \sqrt{\theta (\mathbf{\nu}^2)} \psi
\eqno(98)$$

As before we can use the relation (1) between $d\nu^\ell$ and
$dx^\mu$ to find

$$dx^\mu\Bigl\{ \partial_\mu + \frac{ie}{\hslash c} A_\mu (\textbf{x},
\mathbf{\nu})+ \frac{1}{\chi} \partial_\mu \chi -
\gamma_{h\mu}^\ell (\textbf{x}, \mathbf{\nu})\nu^h
(\partial_\ell^\prime + \frac{1}{\chi}\partial_\ell^\prime \chi)
\Bigl \}\psi = - \frac{i\epsilon mc}{\hslash} \sqrt{\theta
(\mathbf{\nu}^2)} \psi \eqno(99)$$

    By setting $\chi = exp \{\frac{i}{\hslash c} \phi(\textbf{x},
\mathbf{\nu})\}$, this equation (99) can also be written in terms
of nonlinear connections $(N_\mu^\nu$) as follows:

$$dx^\mu\Bigl\{ \partial_\mu + \frac{ie}{\hslash c}( A_\mu (\textbf{x},
\mathbf{\nu})+(1/e)\delta_\mu \phi -(1/e) N_\mu^\ell
\partial_\ell^\prime \phi) - N_\mu^\ell \partial_\ell^\prime
\Bigl\}\psi = - \frac{i\epsilon mc}{\hslash} \sqrt{\theta
(\mathbf{\nu}^2)} \psi \eqno(100)$$

or, in terms of "covariant" $ \frac{\delta}{\delta x^\mu}
=\partial_\mu - N_\mu^\ell \partial_\ell^\prime$, we have

$$dx^\mu\Bigl\{\frac{\delta}{\delta x^\mu} + \frac{ie}{\hslash c}( A_\mu
(\textbf{x},
\mathbf{\nu}) + (1/e) \frac{\delta \phi}{\delta x^\mu})
\Bigl\}\psi = - \frac{i\epsilon mc}{\hslash} \sqrt{\theta
(\mathbf{\nu}^2)} \psi \eqno(101)$$

This equation should be regarded as the "classical" field
equation. Here, $A_\mu(\textbf{x}, \mathbf{\nu}) + (1/e)
\frac{\delta \phi}{\delta x^\mu}$ represents the covariant
four-vector, the external electromagnetic field, in the Finsler
space. Obviously, $A_\mu + (1/e) \frac{\delta \phi}{\delta x^\mu}$
is the gauge transformation of the electromagnetic field and
manifests the gauge invariance of it in this space.

    When the field $A_\mu(\textbf{x}, \mathbf{\nu})$ and the scalar
function $\chi (\textbf{x}, \mathbf{\nu})$re
separable as follows

$$A_\mu(\textbf{x}, \mathbf{\nu})= A_\mu(\textbf{x})-
\gamma_{h\mu}^\ell (\textbf{x}) \nu^h \ddot{A_\ell}
(\mathbf{\nu})\eqno(102)$$

(In fact, we have seen earlier that $\gamma_{h\mu}^\ell$ are
independent of the directional arguments for the Finsler space we
are considering) and

$$\chi(\textbf{x}, \mathbf{\nu}) = \chi(\textbf{x})
\ddot{\chi}(\mathbf{\nu}),\eqno(103)$$

we have

$$dx^\mu \Bigl\{\frac{\delta}{\delta x^\mu} + \frac{ie}{\hslash c}
(A_\mu (\textbf{x}) - N_\mu^\ell \ddot{A_\ell}(\mathbf{\nu}) +
(1/e) \partial_\mu \phi(\textbf{x}) - (1/e)
N_\mu^\ell\partial_\ell^\prime \ddot{\phi}(\mathbf{\nu}))
\Bigl\}\psi $$
$$= - \frac{i\epsilon mc}{\hslash} \sqrt{\theta
(\mathbf{\nu}^2)} \psi \eqno(104a)$$

where $\phi(\textbf{x}) = (\hslash c/i) ln\chi(\textbf{x})$, and
$\ddot{\phi}(\mathbf{\nu}) = (\hslash c/i)
ln\ddot{\chi}(\mathbf{\nu})$, that is,

$$\phi(\textbf{x},\mathbf{\nu})=\phi(\textbf{x}) +
\ddot{\phi}(\mathbf{\nu})\eqno(104b)$$

Also, along the autoparallel curve of the Finsler Space

$$A_\mu(\textbf{x},\mathbf{\nu})dx^\mu = A_\mu(\textbf{x})dx^\mu -
N_\mu^\ell \ddot{A_\ell}(\mathbf{\nu})dx^\mu =
A_\mu(\textbf{x})dx^\mu +
\ddot{A_\ell}(\mathbf{\nu})d\nu^\ell\eqno(105)$$

Now, if we assume

$$\ddot{A_\ell}(\mathbf{\nu}) + (1/e) \partial_\ell^\prime
\ddot{\phi}(\mathbf{\nu})= 0,\eqno(106)$$

then the above field equation takes the following form :

$$dx^\mu \Bigl\{\frac{\delta}{\delta x^\mu} + \frac{ie}{\hslash c}
\bar{A_\mu} (\textbf{x})\Bigl\} \chi =- \frac{i\epsilon
mc}{\hslash} \sqrt{\theta (\mathbf{\nu}^2)} \psi \eqno(107)$$

where

$$\bar{A_\mu} (\textbf{x})=A_\mu(\textbf{x}) + (1/e)\partial_\mu
\phi(\textbf{x})\eqno(108)$$

is the usual gauge transformation in the associated curved space
(Riemannian) of the Finsler space.

    We can now find out quantum field equation starting from (98) and
using (104), (105) and (106) on quantum generalization of the
Finslerian microdomain in two steps as before. The resulting
equation for the bispinor
$\psi_{\alpha\beta}(\textbf{x},\mathbf{\nu})$ for a particle in an
external electromagnetic field $\bar{A_\mu}(\textbf{x})$ is found
to be

$$\gamma_{\alpha\alpha^\prime}^\mu
(\textbf{x})(i\hslash\partial_\mu
-(e/c)\bar{A_\mu}(\textbf{x}))\psi_{\alpha^\prime \beta}
(\textbf{x},\mathbf{\nu})$$
$$- i\hslash\gamma_{\beta\beta^\prime}^\mu (\textbf{x}) \gamma_{h\mu}^\ell
(\textbf{x},\mathbf{\nu}) \nu^h
\partial_\ell^\prime \psi_{\alpha \beta^\prime}
(\textbf{x},\mathbf{\nu}) = mc\psi_{\alpha \beta}
(\textbf{x},\mathbf{\nu})\eqno(109)$$

or, in the following alternative form :

$$\gamma^\mu
(\textbf{x})(i\hslash\partial_\mu
-(e/c)\bar{A_\mu}(\textbf{x}))\psi (\textbf{x},\mathbf{\nu})- \psi
(\textbf{x},\mathbf{\nu})\overleftarrow{\partial_\ell^\prime}(i\hslash
\gamma_{h\mu}^\ell (\textbf{x},\mathbf{\nu}) \nu^h\gamma^{\mu^T}
(\textbf{x})) = mc \psi (\textbf{x},\mathbf{\nu})\eqno(110)$$

The field equation for the antiparticle bispinor field $\psi^c
(\textbf{x},\mathbf{\nu})$ can be obtained from (110) by the
transformation $e \rightarrow - e$. The relation between the
particle and antiparticle fields is found to be

$$\psi^c (\textbf{x},\mathbf{\nu}) =i\gamma^2 \psi^\ast
((\textbf{x},\mathbf{\nu})i\gamma^2\eqno(111)$$

In fact, by taking the complex conjugate of the equation (110) and
then by left as well as right multiplications of it by
$i\gamma^2$, one can arrive at the equation for $\psi^c
(\textbf{x},\mathbf{\nu})$ if the relation (111) is taken into
account.

    If we decompose $\psi (\textbf{x},\mathbf{\nu})$ as in (68), we have

$$\psi^c (\textbf{x},\mathbf{\nu}) = i\gamma^2 (\psi(\textbf{x})
\times\phi^T(\mathbf{\nu}))^\ast i\gamma^2$$
$$ = (i\gamma^2 \psi^\ast(\textbf{x})) \times (i\gamma^2
\phi^\ast(\mathbf{\nu}))^T\eqno(112)$$
        (since $\gamma^{2^T} = \gamma^2$)

Now, if $\psi(\textbf{x})$ represents a particle field in
"$\textbf{x}$-space", then $\psi^c(\textbf{x}) = i\gamma^2
\psi^\ast(\textbf{x})$ is the antiparticle field. Similarly, from
(91) it follows that $i\gamma^2 \phi^\ast(\mathbf{\nu})=
\phi^c(\mathbf{\nu}),$ and therefore it follows from (112) that

$$\psi^c (\textbf{x},\mathbf{\nu}) = \psi^c (\textbf{x}) \times
\phi^{c^T} (\mathbf{\nu})\eqno(113)$$

    For the case of decomposition (59), we can similarly arrive at the
following decomposition of $\psi^c (\textbf{x},\mathbf{\nu})$ :

$$\psi^c (\textbf{x},\mathbf{\nu}) = \psi_1^c (\textbf{x}) \times
\phi^{c^T} (\mathbf{\nu}) + \psi_2^c (\textbf{x}) \times
\phi^{c^T} (\mathbf{\nu})\eqno(114)$$

Here, $\psi_1^c (\textbf{x})$ and $\psi_2^c (\textbf{x})$ are the
eigenstates of $\gamma^0$ with eigenvalues -1 and +1 respectively.

    Now, from the equation (110), by using the decomposition (59) it
is easy to find the following equation for $\psi
(\textbf{x},\mathbf{\nu})$ :

$$(i\hslash\gamma^\mu\partial_\mu - (e/c) \gamma^\mu
\bar{A_\mu}(\textbf{x}) + (3i\hslash b_0/2) \zeta(t)\gamma^0)\psi
(\textbf{x},\mathbf{\nu})$$
$$= (c/e(t))(m+M\zeta (t)e(t))\psi
(\textbf{x},\mathbf{\nu})\eqno(115)$$

Also, by the averaging procedure (64) we can find the field
equation for $\psi(\textbf{x})$ for the "$\textbf{x}$-space" as

$$\Bigl\{i\hslash\gamma^\mu\partial_\mu - (e/c) \gamma^\mu
\bar{A_\mu}(\textbf{x}) + (3i\hslash b_0/2)
\zeta(t)\gamma^0\Bigl\}\psi (\textbf{x})$$
$$= (c/e(t))(m+M\zeta (t)e(t))\psi
(\textbf{x})\eqno(116)$$

The field equation in local inertial frame (the flat Minkowski
space-time) can be derived from (116) with the use of the
vierbeins $V_\mu^a(\textbf{X})$ if one neglects the extremely
small terms as before. The equation is

$$\gamma^\mu(i\hslash\partial_\mu - (e/c)\hat{A_\mu}(\textbf{x}))\psi
(\textbf{x}) = c(m+M\zeta (t)e(t))\psi (\textbf{x})\eqno(117)$$

where $\hat{A_\mu}(\textbf{x}) = e(t)\bar{A_\mu}(\textbf{x})$
(expressed in local coordinates, that is, in the normal
coordinates). Of course, $\hat{A_\mu}(\textbf{x})\approx
\bar{A_\mu}(\textbf{x})$ as $F(t) \approx 1$ in the present epoch
of the universe. The equation (117) is the usual field equation in
Minkowski space-time because the additional time-dependent mass
term $Mc \zeta(t) e(t)$ is negligible in the present era of the
universe, and also it vanishes for the case $Im\{k\} =
Im\{k^\prime\} = 0.$ For non-zero imaginary parts of k and k,
this time-dependent mass term is significant in the very early era
of the universe.

\section{CONCLUDING REMARKS}
\noindent
    The microlocal space-time of extended hadrons is specified here in
accordance with Riemanns original suggestion of a metric function
as the positive fourth root of a fourth order differential form.
The classical field equation has been obtained for this space from
an assumed property of the field on the autoparallel curve, and
the corresponding quantum field equation for a free lepton has
been derived here by making the quantum generalization of this
microlocal Finslerian space-time below a fundamental length-scale.
In this process of quantization, the field transforms into a
bispinor field which can be decomposed as a direct product of the
two spinors depending respectively on the position coordinates and
the directional arguments. The position coordinate-dependent
spinors correspond to the fields for the macroscopic spaces which
are the associated Riemannian space-time of the Finslerian
microdomain. These spinors satisfy the usual Dirac-equations in
those macrospaces. The other spinor depending on the directional
variables satisfies a different equation which has been solved
here. It is shown elsewhere (De, 1997) that these spinors are
responsible for generating additional quantum numbers for the
constituents in the hadron-structure. From this Finsler geometric
origin of internal quantum numbers of the constituents, the
internal symmetry of hadrons was achieved there.
    In the process of separation of the bispinor field and
its equation we have seen that the mass of the particle is $m +
M\zeta(t)e(t)$, where M is related to the parameter bo of the
space-time by (96). Incorporating this relation (96) into the
particle-mass and expressing in terms of the cosmological time T
given in (62) or (63), it is easy to see that the mass of the
particle is $m + \{(0.832\hslash /c^2)H(T)\}$, where the relations
(54) and (57) have also been taken into account. Here $H(T)$ is
the Hubbles function. For a representative particle (muon), the
mass of it is $m(1+2\alpha H(T))$, where $\alpha = 0.26 \times
10^{-23} sec.$ It is evident that the particle-mass at the present
epoch of the universe is approximately equal to its inherent mass
m to an extremely high degree of accuracy. On the contrary, the
epoch-dependent part of the mass had significant contribution
around and before the space time a (after the time-origin of the
universe). In De (1993, 1999, 2001, 2002) its cosmological
consequences in this very early era of the universe have been
discussed. Also, it is to be noted that an "inherent massless"
particle can even have mass $2\alpha \hat{m} H(T)$, $\hat{m}$
being its mass at the epoch $\alpha$.

    The field equations for a particle in an external electromagnetic
field in the Finsler space as well as in the associated Riemannian
spaces have been obtained here. In the subsequent
papers(De,2002a,b) we have considered the electromagnetic
interaction in the Finsler and its associated spaces. The
covariance of the field equations under general coordinate
transformations have been discussed there. Also, the
field-theoretic and the S-matrix approaches for dealing with the
strong interaction of hadrons have been presented.

\pagebreak
\noindent
\underline{\bf{References}}
\begin{flushleft}

Adler, R.J.,and Santiago, D.I. (1999). Modern Physics Letters A,
\textbf{14}, 1371.

Bjorken, J.D. (1969). Physical Review, \textbf{179}, 1547.

Blokhintsev,D.I.(1973). Space and Time in the Microworld, D.
Reidel, Dordrecht, Holland.

De, S.S. (1993). International Journal of Theoretical Physics,
\textbf{32}, 1603.

De, S.S. (1997). International Journal of Theoretical Physics,
\textbf{36}, 89.

De, S.S.(1999).International Journal of Theoretical Physics,
\textbf{38}, 2419.

De, S.S.(2001). International Journal of Theoretical Physics,
\textbf{40}(11), 2067.

De, S.S. (2002). International Journal of Theoretical
                  Physics,\textbf{41}(1), 137.

De S.S. (2002a) International Journal of Theoretical
                  Physics, \textbf{41}, 1291.

De S.S. (2002b) International Journal of Theoretical
                  Physics, \textbf{41}, 1307.

Feynman, R.P. (1972). Photon-Hadron Interactions, Benjamin, New
York.

Gell-Mann, M. (1964). Physics Letters, \textbf{8}, 214.

Hofstadter, R. (ed.) (1964). Electron Scattering and Nuclear and
Nucleon Structure, Benjamin, New York.

Namsrai,Kh. (1985). Nonlocal Quantum Theory and Stochastic Quantum
                     Mechanics, D. Reidel, Dordrecht, Holland.

Ng, Y.J., and Van Dam,H. (1994). Modern Physics Letters A,
\textbf{9}, 335. Prugovecki, E.(1984). Stochastic Quantum
Mechanics and Quantum Space-Time, D.Reidel, Dordrecht, Holland.

Riemann, G.F.B. (1854). Uber die Hypothesen welche der Geometrie
zu Grunde leigen, Habilitation thesis, University of
G\"{o}ttingen.

Rund, H. (1959). The Differential Geometry of Finsler Spaces,
                 Springer, Berlin. Snyder, H.S. (1947). Physical Review,
\textbf{71}, 38.

Yang, C.N. (1947). Physical Review, \textbf{72}, 874.

Yukawa,H.(1948).Progress of Theoretical Physics, \textbf{3},205.

Yukawa,H.(1950). Physical Review, \textbf{77}, 219.

\end{flushleft}
\end{document}